\documentclass[aps,prd,nofootinbib,preprint,tightenlines,longbibliography]{revtex4-1}
\pdfoutput=1

\usepackage[pdftex,breaklinks,colorlinks,
citecolor=blue,
urlcolor=blue]{hyperref}
\usepackage{graphicx}

\newcommand{\be}{\begin{equation}}
\newcommand{\ee}{\end{equation}}
\newcommand{\bea}{\begin{eqnarray}}
\newcommand{\eea}{\end{eqnarray}}

\begin{document}
\title{Eternal inflation, energy conditions and the role of decoherent
  histories}
\author{Eleni-Alexandra Kontou \thanks{Corresponding author}}
\affiliation{ITFA and GRAPPA, Universiteit van Amsterdam,
        Science Park 904, Amsterdam, the Netherlands}
\email{e.a.kontou@uva.nl}
\author{Ken D. Olum}
\affiliation{Institute of Cosmology, Department of Physics and Astronomy, Tufts University, Medford, MA 02155, USA}
\email{kdo@cosmos.phy.tufts.edu }
\date{\today}

\begin{abstract}	
Eternal inflation, the idea that there is always a part of the
universe that is expanding exponentially, is a frequent feature of
inflationary models. It has been argued that eternal inflation
requires the violation of energy conditions, creating doubts for the
validity of such models. We show that eternal inflation is possible
without any energy condition violation, highlighting the important
role of decoherence and the selection of states in the inflationary
process.\\

 \textit{Essay written for the Gravity Research Foundation 2021 Awards for Essays on Gravitation.}

\end{abstract}

\maketitle

Classical physics is deterministic. By specifying the initial
conditions we can exactly determine the correct solution to the
equations of motion. There are quantities such as the total energy
that are conserved by the equations of motion and thus are the same at
all times in each solution.

In quantum mechanics, the closest we can come to this concept is the
set of decoherent histories\footnote{In interpretations without a
  collapse postulate, the evolution of the wavefunction is
  deterministic, however this is not very useful as it includes everything
  that might ever happen.} with corresponding
probabilities. Decoherence is usually understood as interaction with
the environment but in cosmology we must use a notion of effective
decoherence. Here, things are taken to decohere when the information
that could restore their coherence is widely scattered through the
universe. Starting from conditions at a given time, there are many
possible histories that extend these conditions into the future.
Conservation laws are more subtle in this scenario. For example, a
decoherent history might have different expectation values for energy
at different times, even in a model with energy conservation.

As a simple example consider a system with two energy eigenstates
$|E_1\rangle$ and $|E_2\rangle$ and corresponding eigenvalues $E_1$
and $E_2$ that is in state
\be
|\psi\rangle=\epsilon_1 |E_1\rangle+\epsilon_2 |E_2\rangle \,.
\ee
Under unitary evolution, the expectation value $E=\langle \psi |H|\psi
\rangle = |\epsilon_1|^2 E_1+|\epsilon_2|^2 E_2$ does not change, but
this is not the quantity an observer measures.  Upon measurement, an
observer will find the system in energy eigenstate $|E_2\rangle$ with
probability $|\epsilon_2|^2$.  Now the energy is $E_2$ instead of $E$.
This change does not need to be small. If $E_2 \gg E$ the probability
to get $E_2$ is small but not zero. One might ask whether this
change in energy is counterbalanced by a change in energy of the
measurement apparatus or the environment. Interestingly, this question
was recently answered in the negative with the construction of an
explicit counterexample \cite{Carroll:2021ecb}.

Rather than energy conservation, we can consider the restrictions
imposed by energy conditions. For example, the null energy condition
(NEC) guarantees that a converging congruence of null geodesics cannot
be defocused to become diverging. But even in a theory that obeys
NEC, a decoherent history of spacetime may include such a congruence
that converges and then diverges due to the selection of a substate
later in the decoherent history.  This allows eternal inflation to be
possible despite seeming to be ruled out by energy conditions, as we
now discuss.

Energy conditions were introduced early in the history of general
relativity as statements on how ``reasonable'' matter should behave
\cite{Kontou:2020bta}. Here we are mainly concerned with the NEC which
requires the stress-energy tensor $T_{\mu \nu}$ to obey
\be\label{eqn:NEC}
T_{\mu \nu} \ell^\mu \ell^\nu \geq 0 \,,
\ee
for every null vector $\ell^\mu$.  For a perfect fluid stress-energy
tensor, Eq.~(\ref{eqn:NEC}) becomes $\rho+P \geq 0$, where $\rho$ is the
energy density and $P$ the pressure. The NEC is obeyed by most\footnote{The NEC can be violated in some cases by classical non-minimally coupled scalar fields. For more information see \cite{Barcelo:1999hq} and \cite{Kontou:2020bta}.}
classical fields, but it can be
violated by quantum fields as with
all pointwise energy conditions \cite{Epstein:1965zza}. 

\begin{figure}
	\includegraphics[height=5cm]{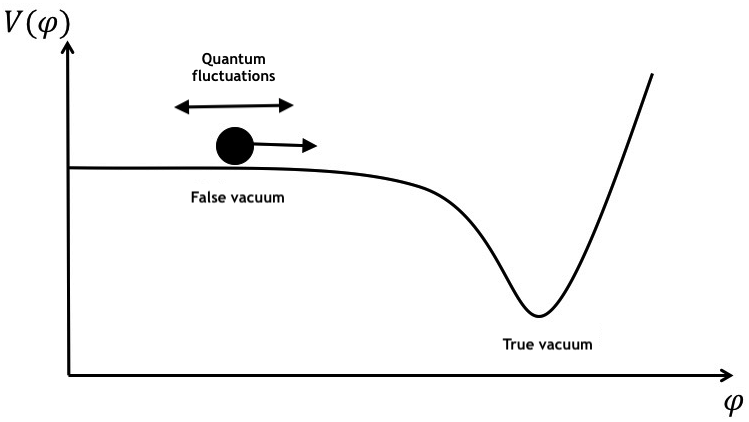}
	\caption{The inflaton classically rolls down the potential slope. However, quantum fluctuations can move it in both directions.}
\label{fig:potential}
\end{figure}

To examine the application of the NEC in cosmology we briefly
introduce the concept of eternal inflation. Inflation is the proposed
period of (near) exponential expansion in the early universe
\cite{Guth:1980zm}. In most inflationary models, this expansion is
driven by a scalar field with a potential, the inflaton. Even though
classically the inflaton rolls slowly down the potential leading to
the end of inflation, its quantum fluctuations can drive the expansion
rate up, as shown in Fig.~\ref{fig:potential}.
While upwards fluctuations are infrequent, because the inflating
volume is exponentially increasing, there are always
parts of the universe inflating in many inflationary models \cite{Guth:2007ng}.

Borde and Vilenkin \cite{Borde:1997pp}, Vachaspati
\cite{Vachaspati:2003de}, and Winitzki \cite{Winitzki:2001fc} argued
that eternal inflation requires violations of the NEC. These works
consider only fluctuations on superhorizon scales, so regions of
horizon-size or a few horizon sizes are approximately given by the
Friedman-Lema\^itre-Robertson-Walker (FLRW) metric,
\be
ds^2=dt^2-a(t)^2 d\vec{x}^2 \,,
\ee
and the equation governing the Hubble expansion, $H=\dot{a}/a$, is
\be
\dot{H}=-4\pi G (\rho+P)+\frac{k}{a^2} \,.
\ee
In an inflating spacetime, we can ignore the curvature term, so set
$k=0$. Eternal inflation requires an increase in the expansion rate,
meaning $\dot{H}>0$, so $\rho+P <0$, and the NEC is violated. 

We will show here that the above argument is incorrect. In fact it is
possible to have eternal inflation even when the NEC is obeyed. This
conclusion is not necessary to show the consistency of eternal inflation
with quantum field theory, because quantum fields can easily violate the 
NEC. However it is obviously sufficient, and once we have made it we no
longer need to be concerned with weaker conditions such as the averaged
null energy condition.  (For details about that condition and how it
interacts with eternal inflation, see \cite{Kontou:2020pdx}.)

How is it possible to have eternal inflation without NEC violation? The NEC tells us that $\dot H \le 0$, but that is a statement about the
evolution of the state of the universe under the equations of motion.
Eternal inflation is not driven by such evolution but rather by the
selection of substates of the quantum state. Indeed, a state with a
given $H$ cannot evolve into one with a larger $H$. Rather, a state
with a given $H$ on average can be composed of substates with
different expansion rates, some of which are larger than the average. Decoherence will select one of the substates, and if it happens to be
one with a high $H$, the resulting decoherent history will have a
larger $H$ at later times than it had before.

For inflation to be eternal, such selections must happen repeatedly. This is possible because more and more modes pass outside the horizon
as the universe expands. We can start with the subhorizon modes in
the Bunch-Davies vacuum. These modes are stretched by the expansion
but while a mode is subhorizon, its frequency changes at a small rate
compared to the frequency itself, thus conditions change adiabatically
and the mode remains in the Bunch-Davies state. For subhorizon
modes, this is close to the Minkowski space ground state and has a
well-defined energy density.

\begin{figure}
	\includegraphics[height=5cm]{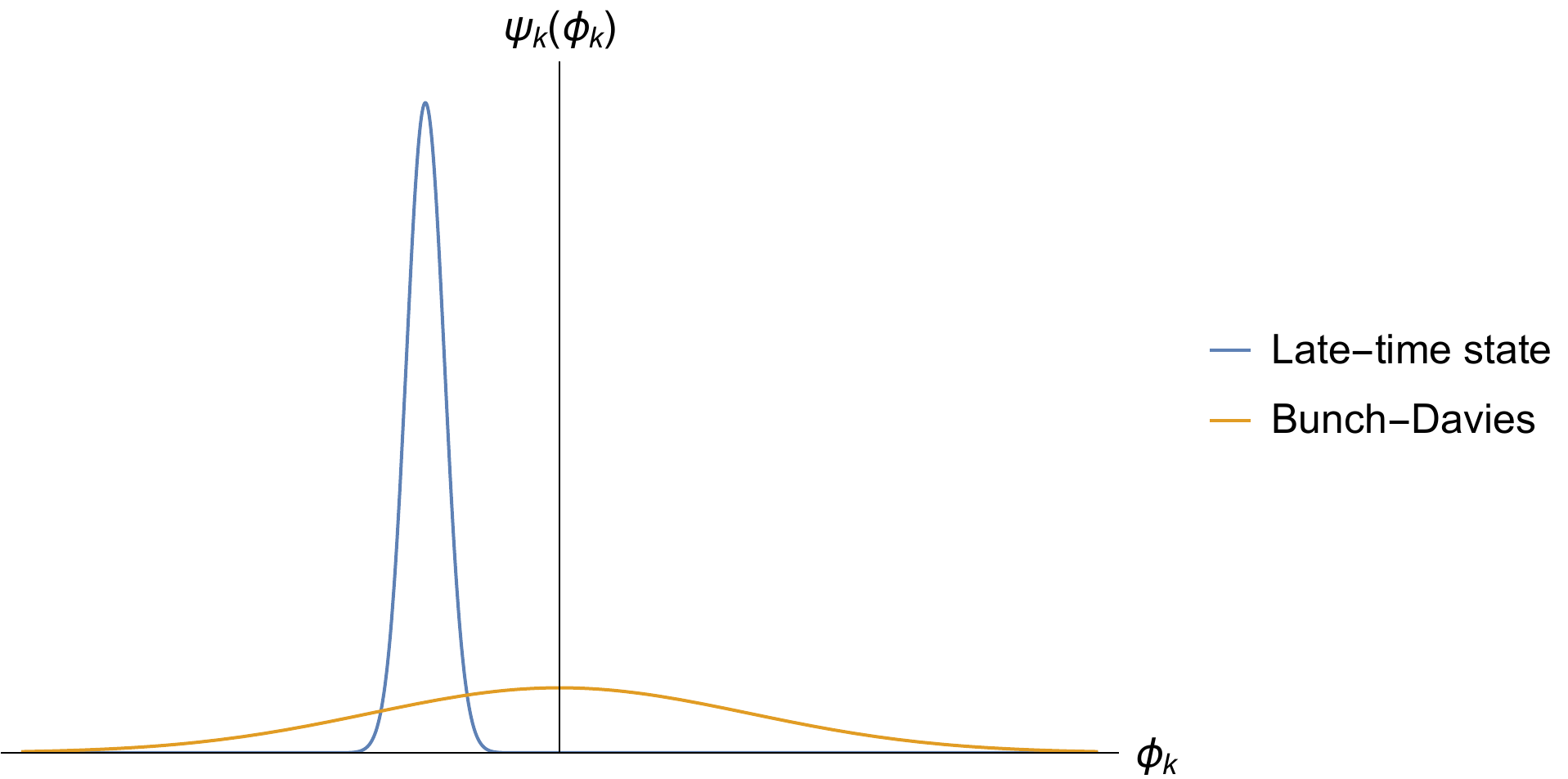}
	\caption{The wavefunction $\psi_k$ for a single mode of the inflaton field. The broad orange
		distribution shows the Bunch-Davies wavefunction for a superhorizon
		mode; the narrow blue distribution shows a possible wavefunction after
		decoherence (late-time state). The latter is peaked around some particular field value $\phi_k$.}
	\label{fig:gaussian}
\end{figure}

However, as the mode crosses the horizon it no longer evolves
adiabatically. The field value is frozen while the momentum is
damped. The field values are now in a wide Gaussian, with different
field values representing different energies (see
Fig.~\ref{fig:gaussian}). Decoherence processes (e.g., see
\cite{Nelson:2016kjm}) select a much narrower state peaked around a
specific field value and thus a specific energy. While the
probability of the state being at a higher energy than the
Bunch-Davies vacuum is small, it is non-zero. Such a state has a
higher expansion rate.

This process occurs repeatedly as modes cross the horizon and
decohere. Roughly each Hubble time there is a new chance for
decoherence to increase $H$. These increases are very rare, but they
result in faster expansion. In most models the expansion is
fast enough that the physical volume of these rare regions is larger
than the much more common ones where $H$ decreases as the inflaton
rolls down the potential hill. Therefore, inflation continues forever.

Thus the NEC is not required to be violated in eternal inflation. Each
state can obey the NEC, while the expansion rate increases through the
selection process. Now we can ask whether typical late-time,
fast-expanding states obey the NEC. We find \cite{Kontou:2020pdx} that
most states leading to eternal inflation do in fact obey NEC. Even though such
states expand rapidly at late times, if we follow them back to earlier
times we find they were expanding even faster. Furthermore, in the
Bunch-Davies vacuum, $T_{\mu \nu} \propto g_{\mu \nu}$, so NEC is
obeyed. The typical states and substates involved in eternal
inflation all obey NEC, even though the transition by decoherence from
one state to another yields an increase in $H$.

In this essay we attempt to clarify how energy conditions should be
applied in cosmology. Importantly, they only apply to
each quantum state individually and we cannot use them to restrict how
a later different state may arise by selection. This observation is
similar to the non-conservation of energy during measurement in a
quantum mechanical system. In cosmology the measuring device is
replaced by decoherence, which can lead to a repeated selection of
states. Therefore an eternally inflating spacetime does not require the
violation of any energy conditions.

\pagebreak

\section*{Acknowledgments}

 We would like to thank Larry Ford, Ben Freivogel, 
 Alan Guth, Tanmay Vachaspati, and Alex Vilenkin for helpful
 conversations. E-AK is supported by the ERC Consolidator Grant QUANTIVIOL. This work is part of the $\Delta$ ITP consortium, a program of the NWO that is funded by the Dutch Ministry of Education, Culture and Science (OCW).

\bibliographystyle{utphys}
\bibliography{paper}

\end{document}